\documentclass[superscriptaddress,showpacs,twocolumn,footinbib,amssymb,aps]{revtex4}

\usepackage{graphicx,dcolumn,bm,amsmath,color,bm}

\newcommand{\eq}[1]{Eq.~(\ref{#1})}

\newcommand{\fig}[1]{Fig.~\ref{#1}}

\newcommand{\eeq}{ \end{equation} }
\newcommand{\beq}{ \begin{equation} }

\newcommand{\eea}{ \end{align} }
\newcommand{\bea}{ \begin{align} }

\newcommand{\bhu}{ {\bf \hat{u}} }

\newcommand{\bq}{ {\bf q} }

\newcommand{\bfr}{ {\bf r} }

\newcommand{\bv}{ {\bf \hat{v}} }
\newcommand{\bw}{ {\bf \hat{w}} }
\newcommand{\bn}{ {\bf \hat{n}} }

\begin{document}

\title{Empty smectics of hard nanorings: insights from a second-virial theory}
\author{H. H. Wensink}
\email{wensink@lps.u-psud.fr}
\affiliation{Laboratoire de Physique des Solides, Universit\'e Paris-Sud  \& CNRS, UMR 8502, 91405 Orsay, France}
\author{C. Avenda\~{n}o}
\affiliation{School of Chemical Engineering and Analytical Science, University of Manchester, Sackville Street, Manchester M13 9PL, UK}

\date{\today}

\begin{abstract}

Inspired by recent simulations on highly open liquid crystalline structures formed by rigid planar nanorings we present a simple theoretical framework explaining the prevalence of smectic over nematic ordering in systems of ring-shaped objects.  The key part of our study is a calculation of the excluded volume of such non-convex particles in the limit of vanishing thickness to diameter ratio. Using a simple stability analysis we then show that dilute systems of ring-shaped particles have a strong propensity to order into smectic structures with an unusual antinematic order while solid disks of the same dimensions exhibit nematic order. Since our model rings have zero internal volume these smectic structures are essential empty, resembling the strongly porous structures found in simulation. We argue that the antinematic intralamellar order of the rings plays an essential role in stabilizing these novel smectic structures.

\end{abstract}

\maketitle
\section{Introduction}

By virtue of their orientation-dependent interactions, non-spherical nanoparticles are capable of displaying a much richer phase morphology than their spherical counterparts. Prominent examples are liquid crystal mesophases which are characterized by broken orientational symmetry (nematic order) combined with long-ranged periodicity in one, two or sometimes three (such as in cholesteric blue phases) spatial dimensions \cite{gennes-prost}. The nature of the simplest liquid crystal, the nematic fluid,  has received a sound statistical-mechanical basis with the classical theory of Onsager \cite{onsager}  in which it is argued that steric repulsion alone can favor nematic states with long-range particle alignment over disordered (isotropic) fluids provided the particle concentration is sufficiently high.  Experimental examples of liquid crystal formation driven by convex non-spherical particle shapes (rods, disks) are quite plentiful in colloid physics \cite{davidson-overview}.  Recent advances in nanoparticle fabrication have led to colloidal or polymeric particles with more complicated,  non-convex shapes \cite{sacanna2013,hernandez_mason2007} with examples ranging from lock-and-key colloids \cite{sacanna_nature2010}, bowl-shaped \cite{marechal_nano2010} and hollow spheres \cite{donath1998} to bent-core \cite{reddy2006} and shape-persistent macrocycles \cite{hoger2000,zhang_moore2006}. Clearly, investigating the spontaneous self-assembly of these intricate particle shapes poses an intriguing challenge to the modelling community \cite{glotzersolomon2007}.

While Onsager-type mean-field theories have been successfully employed to predict the structure and bulk phase behaviour of simple convex bodies (rods, plates, boards, etc.) and their mixtures,  their application to systems of non-convex particle shapes is of more recent date  \cite{blaak1998,blaak2004,dozovEPL2001,Dussi2015a}.  The studies appeared to date have underlined the notion that broken particle symmetry may give rise to intricate periodicity in nematics, involving cubatic \cite{blaak2004} or helical mesostructures \cite{kolli2014b,Dussi2015a}.  A recent computer experiment on assemblies of planar nanorings of different shapes and sizes has revealed striking examples of lamellar order which seems greatly facilitated by the hollow shape of the rings \cite{avendano2009,avendano2016}. Stable smectic structures emerge quite generically provided ample interpenetrability of the rings is guaranteed, i.e., the rings should be sufficiently thin but need not be perfectly round (e.g. regular polygonal rings with at least 4 sides also exhibit smectic order) \cite{avendano2016}. These smectic structures are remarkable for two reasons. First, they are strongly porous since the rings are hollow and therefore have a very small internal volume. This feature is important in view of many materials applications (e.g. the fabrication of photonic crystals) which requires structures with long-ranged periodicity but a low packing fraction as in inverse opals \cite{stein_2008}. The second surprising feature is that the rings are ordered {\em antinematically}, that is, the particle vectors preferentially lie in a plane perpendicular to the nematic director, contrary to what is found in most discotic liquid crystals.  Evidence of antinematicity was found in some soft-interaction models for clay particles \cite{jabbari2014} and deformable dendrimers \cite{georgiou2014}. It is quite surprising to see this type of order emerging in simple systems of nanorings which interact only through steric repulsion without the need to apply an external field \cite{dozov2011}.

In this paper we give a theoretical underpinning for the emergence of empty smectic structures in ring assemblies starting from a simple hard-interaction model treated within a second-virial theory. The approach fully accounts for the non-convex shape of the particles but restricts interactions to pairs only. We show that the theory is capable of reproducing the main features observed in the simulations, namely the prevalence of smectic over nematic order along with a correct assessment of the local antinematic alignment and lamellar spacing. Mixing rings with regular convex disks produces a crossover from  smectic to standard nematic order suggesting that the stability of smectic order must be due to the non-convexity of the rings and their marked propensity to interpenetrate.  We also argue the antinematic or planar nematic order of the rings within the smectic layers to be one of the main contributing factors to smectic stability as it enables the system to retain a much higher degree of orientational entropy than it would if the particles were aligned nematically along  a common director.

\section{Stability of the isotropic fluid against liquid crystalline order}

Without loss of generality we set the thermal energy $k_{B}T =1$ as the unit of energy ($k_{B}$ is Boltzmann's constant and $T$ temperature). The Helmholtz free energy ${\mathcal F}$ of a non-uniform fluid of non-spherical particles is expressed in terms of the one-body density $\rho(\bfr, \bhu)$.  In the second-virial approximation it reads \cite{onsager, allenevans, poni1990, Masters2008}:
\begin{align}
&  {\mathcal F}[\rho] = \mu_{0} +  \int d \bfr d \bhu \rho (\bfr  , \bhu) \ln  [ {\mathcal V} \rho(\bfr, \bhu) - 1 ] \nonumber \\ 
& - \frac{1}{2}   \int d \bfr d \bhu \int d \bfr^{\prime} d \bhu^{\prime} \Phi (\Delta \bfr , \bhu, \bhu^{\prime}) \rho(\bfr, \bhu) \rho(\bfr^{\prime} , \bhu^{\prime}),
\label{free}
\end{align}
with ${\mathcal V}$ the total thermal volume of the particle including contributions from the rotational momenta. The key quantity here is the Mayer function $ \Phi  = e^{-U} -1 $  defined in terms of the orientation-dependent pair potential $U(\Delta \bfr , \bhu, \bhu^{\prime})$ between two non-spherical objects with centre-of-mass distance  $\Delta \bfr = \bfr - \bfr^{\prime}$ and orientation vectors $\bhu$ and $\bhu^{\prime}$.  The chemical potential  $\mu_{0}$ ensures proper normalization of the one-body density, i.e. $\int d \bfr d \bhu \rho(\bfr, \bhu) = N$.

If the particle interactions are strictly hard, which is the case here, then $\Phi = -1$ if the cores overlap and $\Phi = 0$ otherwise. Configurations involving  any number of particle overlaps yield an infinite potential energy and are infinitely improbable. All allowable particle configurations therefore have zero potential energy and the Helmholtz free energy is governed by entropic contributions alone.  This is encoded in \eq{free} where the first term represents the exact translational and orientational entropy of an ensemble of freely rotating non-spherical particles --  both entropies are maximal in a uniform isotropic fluid --  whereas the approximate second contribution accounts for the so-called packing entropy of hard particles by considering only interactions between pairs. Since there are no enthalpic contributions, temperature becomes a mere scaling factor and the overall particle concentration constitutes the only relevant thermodynamic parameter. 

At low particle density the particles will form an isotropic  fluid with uniform particle concentration $\rho_{0} = N/V$ and random orientations so that the one-body density is a simple constant $\rho (\bfr , \bhu) = \rho_{0}/ 4 \pi $. At higher concentration particle-particle interactions will drive liquid crystalline order of some nature.  In order to probe this in more detail we apply a perturbation to the isotropic state by considering an arbitrary density modulation characterized by some wavevector $\bq$ \cite{mulderparns}: 
\beq
\rho (\bfr , \bhu) = \frac{\rho_{0}}{4 \pi}  + \delta \hat{\rho}(\bhu) e^{-i \bq \cdot \bfr}.
\label{rexp}
\eeq
where the amplitude is required to be infinitesimally small, i.e. $ |\delta \hat{\rho}(\bhu) | \ll 1$. This perturbation may signify any type of liquid crystalline order such as nematic, smectic, columnar or even full crystalline order. The perturbation brings about a change in free energy which formally reads up to quadratic order in the amplitude:
\beq
\delta^{2} {\mathcal F}= \int d \bhu d \bhu^{\prime} \left \{ \frac{4 \pi \delta_{ \bhu \bhu^{\prime}}} {\rho_{0}}  - \hat{\Phi}_{ \bq} ( \bhu , \bhu^{\prime} ) \right \} \delta \hat{\rho} ( \bhu)  \delta \hat{\rho} (\bhu^{\prime} ),
\eeq
where $\hat{\Phi}_{\bq}$ represents the Fourier transform (FT) of the Mayer function which for hard interactions reduces to a Fourier transform of the {\em excluded volume} at fixed particle orientation:
\begin{align}
 \hat{\Phi}_{ \bq} ( \bhu , \bhu^{\prime} )    &= \int d \Delta \bfr  \Phi (\Delta \bfr , \bhu , \bhu^{\prime} ) e^{i \bq \cdot \Delta \bfr} \nonumber \\
 & = - \int_{V_{\rm overlap}(\bhu , \bhu^{\prime})} d \Delta \bfr e^{i \bq \cdot \Delta \bfr}.
\end{align}
Local stability of the isotropic fluid against liquid crystalline order requires that $\delta^{2} {\mathcal F}$ be positive, whereas loss of stability happens when $\delta^{2} {\mathcal F} = 0$.  The state-point (particle density $\rho_{0}$) at which this occurs is referred to as a bifurcation point indicating the emergence of liquid crystalline order with a free energy lower than that of the isotropic fluid at the same particle concentration. The bifurcation condition can be established by factorizing the perturbation $\delta \hat{\rho} ( \bhu) = \varepsilon f^{\ast} (\bhu) $ (with $\varepsilon \ll 1$)  and rearranging terms into the following eigenvalue equation \cite{poni1990, roij_mulder1995}:
\beq
f^{\ast} (\bhu)  = \frac{\rho_{0}}{4 \pi} \int d \bhu^{\prime} f^{\ast} (\bhu^{\prime}) \hat{\Phi}_{\bq} ( \bhu , \bhu^{\prime} ),    
\label{bif}
\eeq
where $f^{\ast}(\bhu)$  is the eigenfunction marking the orientational distribution of the particles in the incipient `new' phase.

The stability analysis entails seeking the wave-vector, encoding some prescribed density modulation, that produces the lowest eigenvalue  $\rho_{0}$. The latter is identified as the bifurcation density.  The simplest liquid crystalline instability is the nematic. This state is characterized by a uniform density but a non-uniform orientation probability in which the particles adopt a certain degree of alignment along a common nematic director denoted by $\bn$.  The bifurcation towards the nematic is particularly straightforward to gauge since there is no periodicity (i.e., $\bq \rightarrow 0$) while the incipient orientation distribution takes on the form of a simple Legendre polynomial $f^{\ast} (\bhu \cdot \bn) = {\mathcal P}_{2}(\bhu \cdot \bn)/4 \pi $ (with ${\mathcal P}_{2}(x) = \frac{3}{2} \cos^{2} x - \frac{1}{2}$) \cite{kayser}. The isotropic-nematic bifurcation density then simply follows from \eq{bif} after some basic rearrangements:
\beq
\rho_{0}^{\ast} = \frac{\langle ({\mathcal P}_{2}(t))^{2} \rangle_{t} }{\langle \langle {\mathcal P}_{2}(t) {\mathcal P}_{2}(t^{\prime}) \langle B_{2} (\bhu , \bhu^{\prime} ) \rangle_{\Delta \varphi}  \rangle_{t} \rangle_{t^{\prime}} },
\label{isonem}
\eeq   
with $t = \bhu \cdot \bn$ the projection of the particle vector onto the nematic director and $\Delta \varphi$ the azimuthal angles describing the relative particle orientation in the plane perpendicular to the director so that we may parameterize $(\bhu, \bhu^{\prime}) \rightarrow (t , t^{\prime}, \Delta \varphi)$. 
The brackets denote averages over the polar projections, $\langle \cdot \rangle_{t}  = \int_{-1}^{1} dt$, and azimuthal orientations $\langle \cdot \rangle_{\Delta \varphi} = (2 \pi)^{-1} \int_{0}^{2 \pi} d \Delta \varphi$. The key ingredient here is the {\em second-virial coefficient} $B_{2}(\bhu , \bhu^{\prime} )  = \frac{1}{2} \hat{\Phi}_{0} (\bhu, \bhu^{\prime} ) $ defined  as the excluded volume per particle \cite{onsager}. For slender uniaxial particles (needles, disks, rings) this quantity is strongly orientation-dependent and scales as  $B_{2}(\bhu , \bhu^{\prime} ) \propto |\sin \gamma |$   in terms of the enclosed angle $\gamma$ between the main particle vectors. The corresponding bifurcation density then simply follows from $\rho_{0}^{\ast} = 4 / \langle \langle B_{2} \rangle \rangle_{\rm iso} $ in terms of isotropic average of the second virial coefficient \cite{kayser}.

\begin{figure}
\begin{center}
\includegraphics[width= \columnwidth]{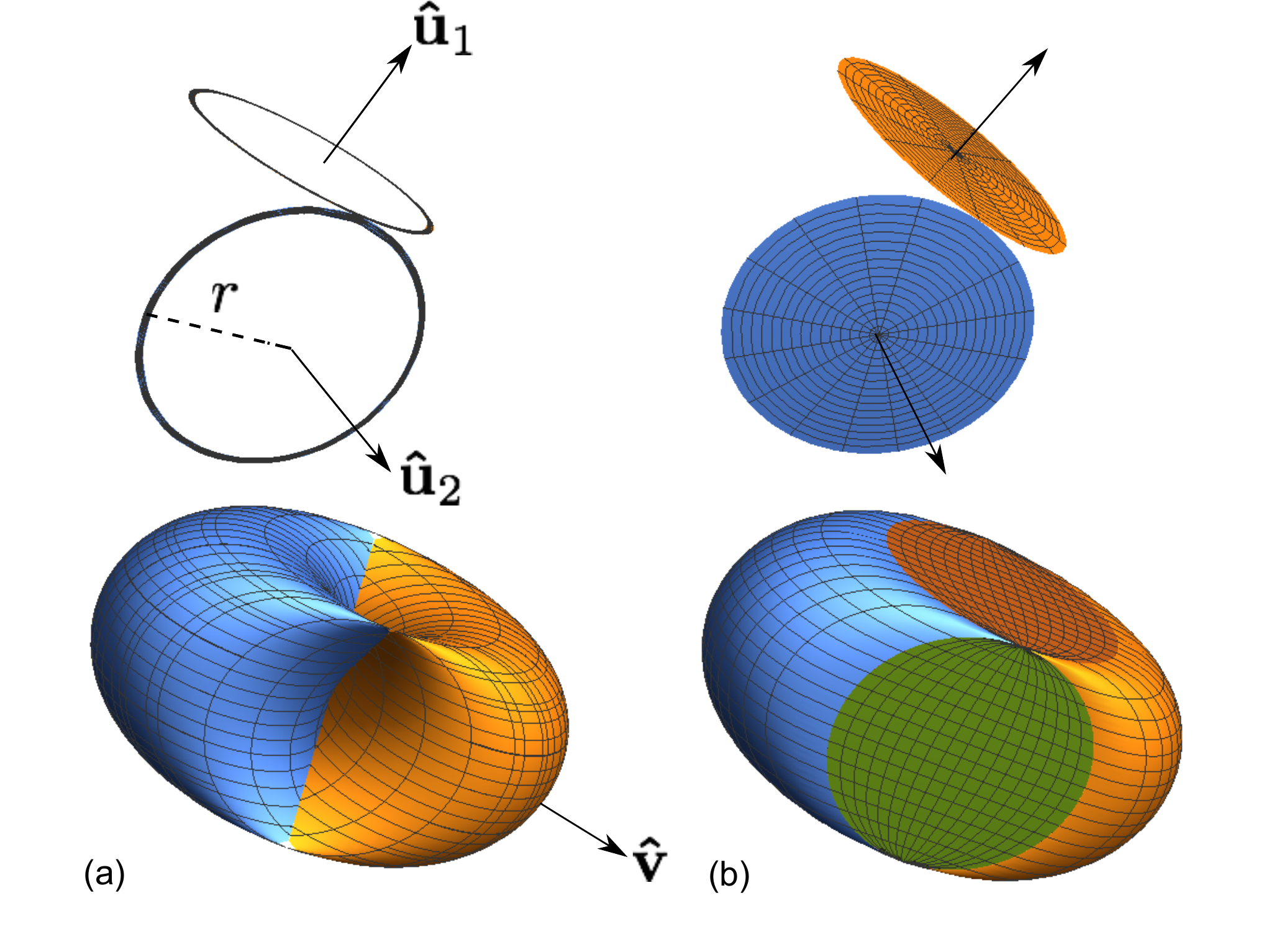}
\caption{ \label{rings} Visualization of the excluded volume of a pair of infinitely thin hard rings (a) and disks (b) with radius $r=D/2$ at fixed mutual orientation defined within a particle-based frame spanned by the three unit vectors. The overlap figure of rings is strongly non-convex and contains sharp inward cusps. }
\end{center}
\end{figure}

\section{Excluded volume of rigid rings and disks}

\begin{figure}
\begin{center}
\includegraphics[width= 0.8 \columnwidth]{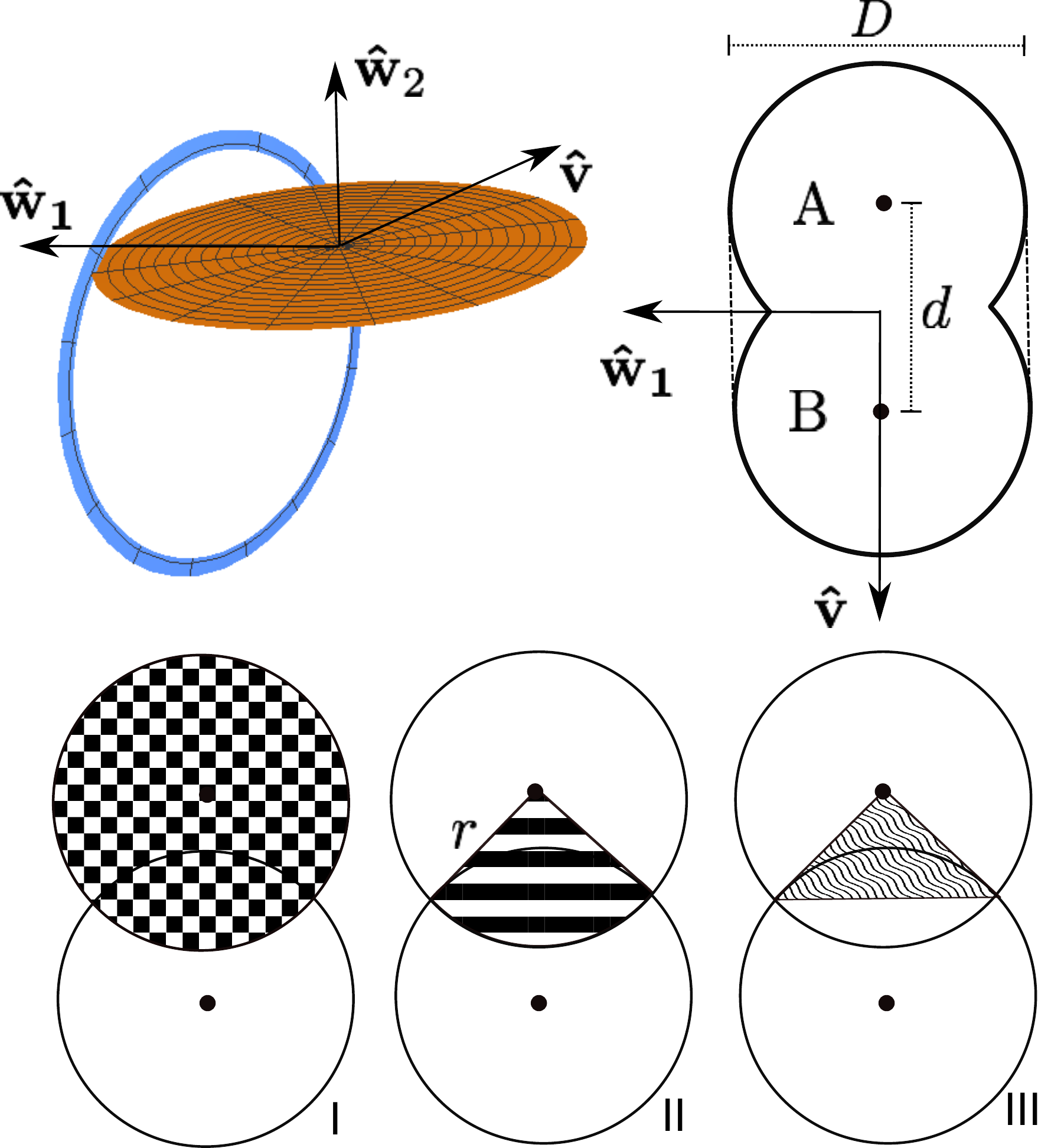}
\caption{ \label{overlap} Overlap between a ring and a disk;   rolling the disk around the circle in the $\{ {\bf \hat{w}_{1}}, {\bf \hat{v}} \}$-plane at fixed mutual orientation projects a typical dimer composed of two overlapping circles with radius $r = D/2$. Its surface can be calculated by decomposing the area into a circular section (I, checkerboard), a circle segment (II, stripes), and a triangular section (III, waves). The dotted lines in the top-right sketch denotes the convex hull of the fused circles.}
\end{center}
\end{figure}

The key quantity of interest here is the excluded volume between the two hard objects at fixed mutual orientation defined as the figure swept out as one object moves around the other at closest contact.  The excluded volume of spherical particles is simply another spherical object with twice the radius of its constituents.  Anisotropic objects, however, produce much more complicated geometries depending on their mutual orientation. \fig{rings}  depicts  the excluded volume manifolds for the particles under scrutiny; hollow rings and filled disks. Both objects are characterized by a diameter $D=2r$ and a thickness $L$ which is assumed asymptotically small so that $L/D \downarrow 0$. The rings are infinitely thin and are {\em not allowed to interlock}. Thus, both objects have an internal volume tending to zero, but a {\em finite} excluded-volume which is non-trivially orientation-dependent. Clearly, the figure associated with the excluded-volume zone of two rings is highly-non-convex due to the interpenetrability of the particles. This gives rise to sharp cusps located at the four square edges of the figure which  join together at the centre-of-mass of the body (\fig{rings}), reflecting the possibility of a complete overlap of mathematical rings at mutual perpendicular orientation $\bhu_{1} \cdot \bhu_{2} = 0$, a configuration resembling the gimbals of a gyroscope. Complete overlap of disks is not possible unless the particles are strictly parallel ($\bhu_{1} \cdot \bhu_{2} \rightarrow 1$) in which case the overlap volume vanishes.  The main challenge confronting us is to parameterize the non-convex overlap manifold associated with the rings. While several routes are conceivable, we find that the most expedient one involves computing the overlap between a ring and a disk as sketched in \fig{overlap}. Let us first define a particle-based frame from the normal vectors $\bhu_{1}$ and $\bhu_{2}$ of a pair of disks or rings. Defining the additional unit vectors $\bv = \bhu_{1} \times \bhu_{2}/ |\bhu_{1} \times \bhu_{2} | $ and $\bw_{i} = \bhu_{i} \times \bv$ so  that $\{ \bhu_{i} , \bv , \bw_{i} \} $ (with $i=1,2$) we obtain  two orthonormal frames. The excluded volume is most conveniently parameterized in the {\em non}-orthogonal, rhombic $\{ \bw_{1} , \bv , \bw_{2} \} $  frame with unit volume $|\bhu_{1} \times \bhu_{2} | = |\sin \gamma |$.

First, we parameterize the circular parts (I) as follows
\begin{align}
\bfr_{{\rm I}} ^{\rm A}&= -d\bv + t_{1} \sin \xi \bv   + t_{1} \cos \xi \bw_{1} + t_{3}  \bw_{2}  \nonumber \\
\bfr_{{\rm I}}^{\rm B} &= d\bv - t_{1} \sin \xi \bv   + t_{1} \cos \xi \bw_{1} + t_{3}  \bw_{2},  
\end{align}
with integration limits $0 < t_{1} < r$, $-r < t_{3} < r$ and $0 < \xi < 2 \pi$ and $d = (r^{2} - t_{3}^{2})^{\frac{1}{2}}$ the centre-to-centre distance of the fused circles.
For the circle segments (II) we use the same form as above but with the angular integration replaced by $-\cos \frac{d}{r} < \xi < \cos \frac{d}{r}$.
Finally, the triangular parts (III) can be parameterized via
\begin{align}
\bfr_{{\rm III}}^{\rm A} &= -t_{1} \bv + t_{2} \bw_{1} + t_{3} \bw_{2}  \nonumber \\
\bfr_{{\rm III}}^{\rm B} &= t_{1} \bv + t_{2} \bw_{1} + t_{3} \bw_{2},  
\end{align}
 with $0 < t_{1} < d$, $-(t_{3} - t_{1} \tan \frac{d}{r}) < t_{2} < (t_{3} - t_{1}\tan \frac{d}{r})$, $-r < t_{3} < r$.  The FT of the excluded volume per particle (i.e. the second virial coefficient) for a ring-disk (RD) pair is given by a linear combination of the three contributions via
 \begin{align}
\hat{B}_{2\bq}^{{\rm RD}} &=  \sum_{\rm A,B} \left [ \int d \bfr_{{\rm I}} e^{i \bq \cdot \bfr_{{\rm I}}}  -  \int d \bfr_{{\rm II}} e^{i \bq \cdot \bfr_{{\rm II}}} + \int d \bfr_{{\rm III}} e^{ i \bq \cdot \bfr_{{\rm III}}} \right ].
 \label{contour}
\end{align}
The integrals can be worked out by invoking the coordinate transformations $\int d \bfr_{i} \rightarrow \int dt_{1} \int dt_{3} \int d \xi J_{i}$ ($i={\rm I}, {\rm II}$) and $\int d \bfr_{\rm III} \rightarrow \int dt_{1} \int dt_{2} \int d t_{3} J_{\rm III}$ with $J_{i}$ being the Jacobian matrix associated with the transformation.   While the results for arbitrary non-zero wavevector cannot be obtained in closed form, the actual excluded volume can be retrieved analytically from the zero wavenumber limit $\hat{B}^{\rm RD}_{2\bq \downarrow 0} = B_{2}^{\rm RD}$ which yields
\beq
B_{2}^{\rm RD} = D^{3} \left ( \frac{1}{3} +  \frac{\pi}{8} \right ) | \sin \gamma |. 
\eeq 
The FT of the second-virial coefficient between two rings (RR) is now easily obtained from
\beq
\hat{B}_{2 \bq}^{{\rm RR}}  = \hat{B}_{2\bq}^{{\rm RD}} +  \hat{B}_{2\bq}^{{\rm DR}}  -  \hat{B}_{2\bq}^{{\rm DD}}. 
\label{b2qrr}
\eeq
The contribution for disks (DD) in Fourier space has been derived in Ref. \cite{wensink_trizac14} and can be readily reconstructed from \fig{overlap} by considering the convex hull of the dimer area (no cusps, see dotted lines in \fig{overlap}) resembling a 2D spherocylinder.  Also here, the zero wavelength limit is well-known and yields the excluded-volume between two infinitely thin hard disks with diameter $D$, namely $B_{2}^{\rm DD} = \frac{\pi }{4} D^{3}  | \sin \gamma |  $. The result for two rings then follows from \eq{b2qrr} and turns out
\beq
B_{2}^{\rm RR} = \frac{2}{3}D^{3}  | \sin \gamma |. 
\eeq 
The ratio  $ B_{2}^{\rm RR} / B_{2}^{\rm DD} = 8/3 \pi \approx  0.85$ indicating that the excluded volume of hollow rings is only about 15 \% smaller than that of disks providing the particle pairs have the same diameter and mutual orientation.
The finite wavenumber values of the second-virial coefficients were obtained by numerically solving the contour integrals in \eq{contour} using standard numerical integration packages \cite{ram2010}. These values then numerically define the kernel of the eigenvalue equation \eq{bif} which is subsequently solved using a matrix diagonalization routine by discretizing the orientational phase space in terms of a discrete number of polar ($ 0< \theta < \pi $) and azimuthal  ($0< \varphi < 2 \pi$) angles with respect to the nematic director.  Note that the polar angle, measuring the projection of the particle normal onto the nematic director via $\cos \theta = \bhu \cdot \bn $, is the only relevant angle for describing uniaxial nematic order we consider here.

\section{Results for a binary mixture of rings and disks}

We now have all the ingredients to investigate the various instabilities that might occur in the isotropic fluid. In order to smoothly interpolate between the convex disk and non-convex ring shape we will consider a binary mixture of the two. Let us define $x$ as the mole fraction of rings, then the FT of the second virial coefficient of the mixture can be approximated by:
\beq
\hat{B}_{2\bq} ^{\rm mix} = (1-x)^{2} \hat{B}_{2\bq}^{{\rm DD}}   + (1-x)x  \hat{B}_{2\bq}^{{\rm DR}}  + x(1-x) \hat{B}_{2\bq}^{{\rm RD}}   + x^{2} \hat{B}_{2\bq}^{{\rm RR}}.  
\eeq
We stress that this form is a simplified one; it presupposes that both species undergo the same spatial density modulation and that there are no compositional fluctuations contributing to the loss of stability of the isotropic fluid. A more elaborate treatment allowing for a full coupling between orientational, positional and compositional degrees of freedom is realizable but goes beyond the scope of the present work.   The isotropic-nematic instability (I-N) is the easiest to establish from \eq{isonem}. Given that the second-virial coefficients of rings and disks only differ by a constant prefactor we can immediately deduce that the I-N instability of pure rings ($x=1$) should occur at a density which is a factor $3 \pi / 8 \approx 1.17$ higher than that of the disks. Of course, we need to keep in mind that the nematic phase need not be the first stable phase as transitions to smectic or columnar order might pre-empt it.  The smectic A  (SmA) phase is characterized by a unidirectional density modulation along the nematic director whereas columnar order implies two-dimensional ordering in the plane perpendicular to $\bn$. We thus decompose $\bq = {\mathcal D} \cdot\bn$ with ${\mathcal D} = q_{\parallel} \bn \bn  + q_{\perp} ({\bf I} - \bn \bn)$ such that ($q_{\parallel} >0$, $q_{\perp} = 0$) imposes smectic order and ($q_{\parallel} =0$, $q_{\perp} > 0$) columnar order. 
\begin{figure}
\begin{center}
\includegraphics[width=  0.8 \columnwidth]{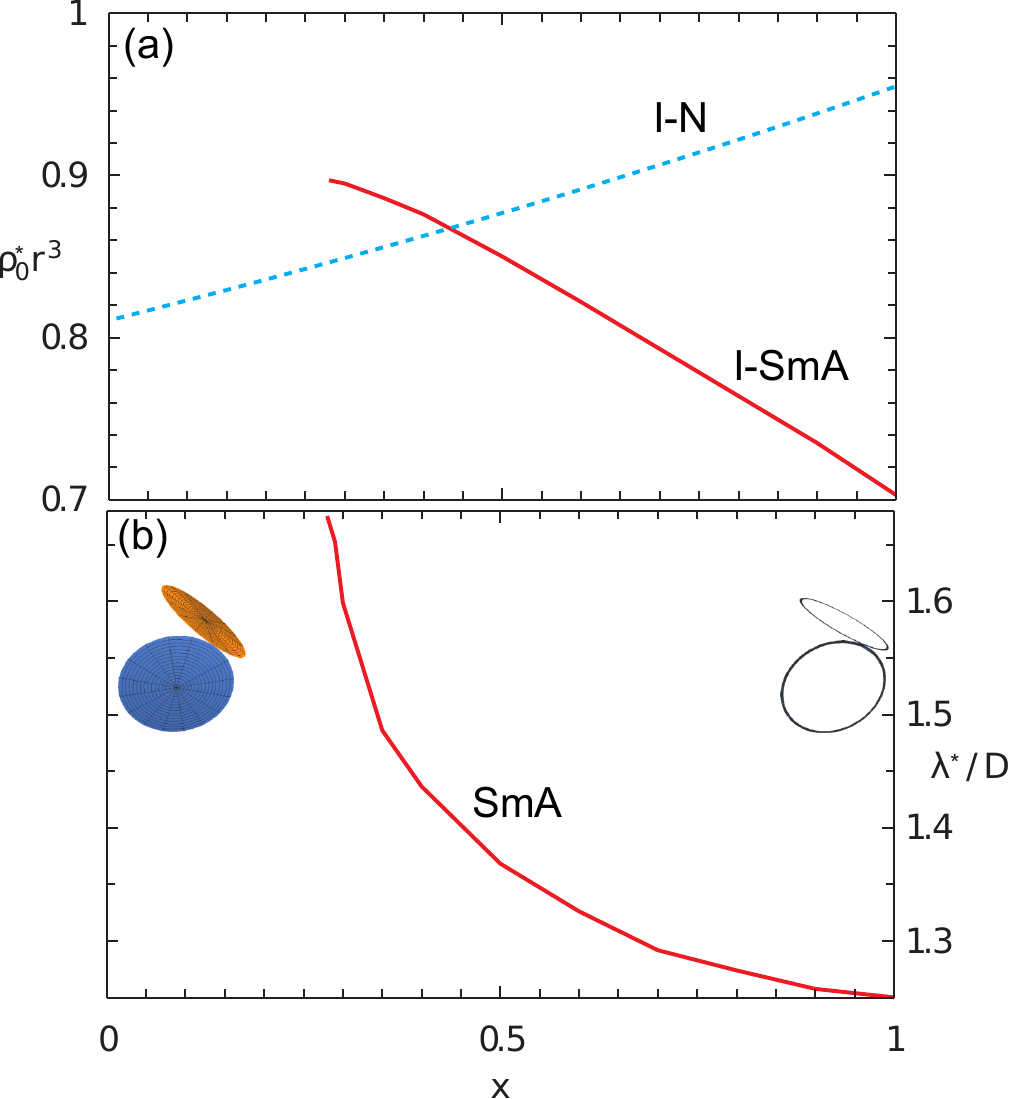}
\caption{ \label{fig2} (a) Bifurcations in a binary isotropic fluid mixture of hard nanodisks and rings. Shown is the normalized particle density $\rho^{\ast}_{0}r^{3}$ on the vertical axis versus the ring mole fraction $x$. The emergent type of liquid crystalline order is given by the curve with the lowest density. Pure rings ($x=1$) exhibit a direct transition from isotropic to smectic order whereas pure systems of disks ($x=0$) form a nematic phase.  (b) Characteristic lamellar distance $\lambda^{\ast}$ corresponding to the smectic phase expressed in units of the particle diameter $D$. }
\end{center}
\end{figure}

\fig{fig2} provides an overview of the main results. The two principal instability modes, the isotropic-nematic (I-N) and the isotropic-smectic (I-SmA) one, are shown in \fig{fig2}a.  Naturally, the one with the lowest density represents the physically relevant instability as it indicates the first new phase appearing upon densification of the isotropic fluid.  While an isotropic fluid of pure disks ($x=0$)  becomes nematic, a dilute system of rings shows a clear tendency to form smectic phases at higher concentration without the intervention of nematic order. This is in agreement with what has been observed in the simulations \cite{avendano2016}. The isotropic-smectic bifurcation is located at $\rho_{0}^{\ast}r^{3} \approx 0.7$  pre-empting the I-N one ($\rho_{0}^{\ast} r^{3} = 3/\pi \approx 0.95$) by more than 25 \%.  
Irrespective of  composition, we found that the isotropic-columnar (I-Col) bifurcation (results not shown) is always located at densities well above the other curves so columnar order does not interfere with the other modes even though the I-Col bifurcation density shifts to considerably smaller values going from pure disks to pure rings.
We reiterate that the smectic structures predicted by our analysis are essentially empty because of the following: i) the transition takes place at finite particle concentration, and ii) the rings have a vanishing internal volume. This scenario is in stark contrast with columnar phases emerging from dense nematic systems of infinitely thin disks. Here,  even though the internal volume of the disks vanishes upon reducing the aspect ratio $L/D \downarrow 0$, the critical particle concentration at which the N-Col transition occurs diverges in such  a way that the product of the two quantities, the packing fraction, always attains a finite value of around 40 to 45 \% \cite{bates_frenkel1998,marechal2011}. 
Upon increasing the mole fraction of disks ($x<1$) the I-N transition exhibits a shallow downward trend reflecting the very similar excluded volumes of the rings and disks (their prefactors differ only by 15 \%). The I-SmA instability, however, abruptly terminates below some critical mole fraction of disks. This indicates a complete absence of the smectic mode for the pure disks at least coming from the isotropic phase. The disruptive effect of the disks on the smectic structure is also reflected in the lamellar spacing  [\fig{fig2}b] which rapidly grows up to almost twice the ring diameter upon increasing the fraction of disks. These large spacings are unlikely to occur spontaneously and it is conceivable that equimolar ring-disk mixtures are prone to form segregated binary smectic structures in which each component obeys a different smectic periodicity and/or internal orientational order.  The discussion of this interesting problem is beyond the scope of the present work and will be discussed in a future paper.

\begin{figure}
\begin{center}
\includegraphics[width= 0.8 \columnwidth]{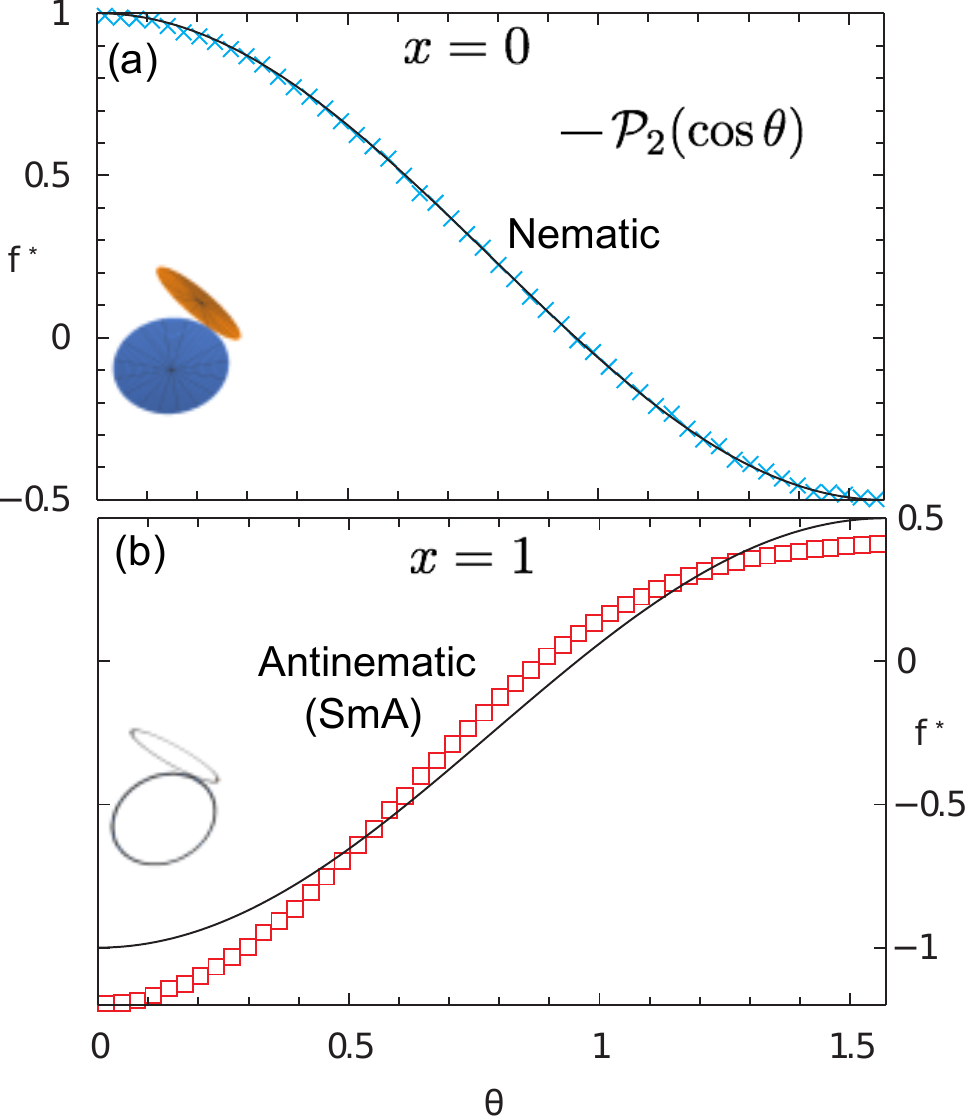}
\caption{ \label{fig3}  Characteristic eigenfunctions  $f^{\ast}$ (from \eq{bif}) indicating the preferred orientational order in the new phase: (a) nematic phases of pure disks show regular uniaxial nematic order, whereas smectic phases composed of rings exhibit typical {\em antinematic} order (b). The orientational distribution in the antinematic smectic phase deviates from the standard second-order Legendre polynomial form (black curve).}
\end{center}
\end{figure}

The eigenfunctions provide information about the orientational order the particles adopted by the emerging phase.  Examples for the pure systems are shown \fig{fig3}. As expected, the nematic phase of disks clearly exhibits ${\mathcal P}_{2}$ type order with the disk normals pointing on average along the nematic director. The rings, on the other hand, are characterized by a typical {\em antinematic} order in which the ring normals preferentially point perpendicular to the nematic director (hence the peak at $\theta = \pi/2$).  This is in complete accordance with the structures that have been established in the simulation model \cite{avendano2016}.   In view of the normalization of $\rho(\bfr,  \bhu)$ [\eq{rexp}]  the eigenfunctions must obey $\int_{0}^{1} d (\cos \theta) f^{\ast}(\theta ) =0$. While the incipient nematic order of the disks follows the typical ${\mathcal P}_{2}$ form [\fig{fig3}a], the antinematic order of the rings cannot be fitted to such a form [\fig{fig3}b]. The nematic order parameter is given by $S \propto \varepsilon \int_{0}^{1} d (\cos \theta) f^{\ast} (\theta) {\mathcal P}_{2} (\cos \theta) $ ($\varepsilon >0$) and yields $S = \varepsilon/5$ for the disks \cite{stroobants1984} and a very similar but negative value for the rings, $S \approx -0.2 \varepsilon$ (note that perfect antinematic order gives $S=-0.5$).

We now wish to briefly illustrate the role of antinematic order in the stabilization of smectic structures by analyzing the orientational entropy. Let us assume that the normalized orientational probability density in a strongly ordered regular uniaxial nematic phase can be described by a simple Gaussian, $f_{\rm N}(\theta)  \sim (\alpha  / 4\pi ) \exp (-\alpha \theta^{2}/2)$, complemented with its mirror form $f_{\rm N}(\pi -\theta)$ for the probability density anti-parallel to the director \cite{Vroege92}.  The variational parameter $\alpha$ is proportional to the nematic order parameter and we require $\alpha \gg 1$. The distribution in the antinematic (AN) phase is peaked around the perpendicular polar angle $\theta = \frac{\pi}{2}$ and the Gaussian distribution reads  $f_{\rm AN}(\theta)  \sim \sqrt{\alpha / (2 \pi )^{3}} \exp (-\alpha (\frac{\pi}{2} - \theta)^{2}/2)$ in normalized form \cite{wensinkrodplate}. The orientational entropy per particle associated with these distributions is defined as $\sigma^{\rm or}  \propto -k_{B} \int d \bhu  f(\bhu)  \ln [4 \pi f(\bhu) ] $ and yields $\sigma_{\rm I}^{\rm or} = 0$ for the isotropic phase, $\sigma^{\rm or}_{\rm N} \sim - k_{B} \ln \alpha $ for the nematic and $\sigma^{\rm or}_{\rm AN} \sim - \frac{1}{2} k_{B}  \ln \alpha $ for the antinematic phase up to leading order in $\alpha \gg 1$.  From this observation we infer that the orientational entropy of the antinematic phase is much higher than that of the nematic phase at least in the limit of asymptotically strong alignment.  This provides a clue as to why smectic order might be preferred over nematic order.  The antinematic organization of the rings is primarily  driven by the additional free volume that is generated when the ring centres-of-mass are co-planar and their normal vectors are mutually perpendicular as observed in the simulations \cite{avendano2016}. In this configuration the rings are allowed to interpenetrate completely ({\em cf.} the cusp in  \fig{rings}a). A similar reduction of pair excluded volume could have been achieved by  a simple nematic alignment of the ring normals but the associated orientational entropy would be much smaller. Antinematic smectic order then may become thermodynamically favorable over simple nematic alignment if the gain in orientational entropy outweighs the reduction of the translational entropy of the smectic associated with its lamellar structure. 
 
\section{Conclusions}

Inspired by recent simulation evidence of novel porous lamellar structures formed in assemblies of nano-rings, we have proposed a simple second-virial route to investigating the onset of liquid crystal order in systems of hard ring- and disk-shaped objects, as well as mixtures of both species.  Our main finding is that a simple, non-ideal fluid description based on the virial expansion, originally devised for regular convex bodies, is also capable of predicting the salient features of liquid crystalline order in assemblies non-convex, hollow particles. Our  simple second-virial theory predicts the emergence of smectic phases at finite particle concentration along with the typical lamellar spacing as well as an antinematic intralamellar orientational order. Since our model is based on mathematical rings with no internal volume, the packing fraction of these smectic phases is essentially zero, indicating empty structures.  We have rationalized the stability of these smectics (with respect to regular nematic phases) from the favorable orientational entropy associated with the antinematic orientational order of the rings within the smectic layers.

\section*{Acknowledgments}

The authors would like to thank George Jackson for inspiring discussions and for stimulating our interest in this problem. 

\bibliography{refs}

\end{document}